\documentclass[referee]{aa}  

\usepackage{longtable}
\usepackage{graphicx}
\usepackage{txfonts}

\usepackage{natbib,twoopt}
\usepackage[breaklinks=false]{hyperref} 
\bibpunct{(}{)}{;}{a}{}{,}             

\newcommand{\dg}{$^\circ$} 
\newcommand{\EV}{EV$_5$} 
\newcommand{\SIu}{J\,m$^{-2}$s$^{-1/2}$K$^{-1}$}

\begin{document} 
\title{Thermophysical properties of near-Earth asteroid \\ (341843) 2008 EV$_5$ from WISE data}

\author{
  V. Al\'i-Lagoa
  \inst{1,2}
  \and
  L. Lionni
  \inst{3}
  \and 
  M. Delbo  
  \inst{4}
  \and 
  B. Gundlach
  \inst{5}
  \and 
  J. Blum
  \inst{5}
  \and
  J. Licandro
  \inst{1,2}
}

\institute{
  Instituto de Astrof\'isica de Canarias (IAC), 
  c/ V\'ia L\'actea s/n, 38205, La Laguna, Tenerife, Spain
  \\
  \email{vali@iac.es}
  \and
  Departamento de Astrof\'isica, Universidad de La Laguna. 38206, La Laguna, Tenerife, Spain
  \and
  University Paris VII - Diderot, 5 Rue Thomas Mann, 75013 Paris, France
  \and
  UNS-CNRS-Observatoire de la C\^ote d'Azur, B.P. 4229, 06304 Nice Cedex 4, France
  \and 
  Institut f\"ur Geophysik und extraterrestrische Physik, Technische Universit\"at Braunschweig, Mendelssohnstr. 3, D-38106 Braunschweig, Germany
}

\date{
  Received, 2013; accepted for publication in Astronomy \& Astrophysics, 2013}

\abstract
    {}
    {
      To derive the thermal inertia of 2008 \EV, the baseline target for the Marco Polo-R mission proposal, and infer information about the size of the particles on its surface. 
    }
    {
      Values of thermal inertia are obtained by fitting an asteroid thermophysical model to NASA's Wide-field Infrared Survey Explorer (WISE) infrared data. 
      From the constrained thermal inertia and a model of heat conductivity that accounts for  different values of the packing fraction (a measure of the degree of compaction of the regolith particles), grain size is derived. 
    }
    {
      We obtain an effective diameter $D = 370 \pm 6\,\mathrm{m}$, geometric visible albedo $p_V = 0.13 \pm 0.05$ (assuming $H=20.0 \pm 0.4$), and thermal inertia $\Gamma = 450 \pm 60$ \SIu at the 1-$\sigma$ level of significance for its retrograde spin pole solution. 
      The regolith particles radius is $r = 6.6^{+1.3}_{-1.3}$ mm for low degrees of compaction, and $r = 12.5^{+2.7}_{-2.6}$ mm for the highest packing densities.
    }
    {}
    

\titlerunning{Thermophys. prop. of 2008 EV$_5$ from WISE data}

\maketitle
%

\section{Introduction}
\label{sec:intro} 

Potentially hazardous, near-Earth asteroid (341843) 2008 \EV, hereafter \EV, has been selected as the baseline target of the sample return mission Marco Polo-R, proposed to the European Space Agency with launch window between 2020 and 2024 (see https://www.oca.eu/MarcoPolo-R/index.html). 
Studying the nature of the surface of \EV\, is therefore important, because collecting samples to obtain unaltered material will require different technological approaches depending on whether the outer layer consists of bare rock, fine-grained, or coarse-grained regolith. 
Information about regolith grain size can be derived using the heat conductivity model of \citet{Gundlach2013} given a value of the thermal inertia, which in turn can be constrained by fitting a thermophysical model (TPM) to infrared data \citep[see, e.g.,][]{Spencer1989,Spencer1990,Lagerros1996I}. Previous thermophysical studies of other mission targets can be found in, e.g., \citet{Mueller2005,Mueller2011,Wolters2011,Mueller2012}.

In addition, deriving surface physical properties of \EV\, is interesting per se. 
First, thermal inertia plays a key role in the Yarkovsky effect, a non-gravitational dynamical force that induces a gradual drift in the orbits of asteroids with sizes of the order of 40 km and smaller \citep{Bottke2006}. 
Accounting for this effect is essential to determine accurate orbits of these objects, especially those classified as potentially hazardous \citep[see][and references therein]{Delbo2009}. 
Second, it was a very strong radar target that produced a high-resolution shape model at its December 2008 Earth approach \citep{Busch2011}. 
Finally, because in spite of the fact that its visible-to-near-infrared spectrum suggests that \EV\, belongs to the C-complex \citep{Reddy2012ev5} and is thus carbonaceous-rich, its diameter of 400 $\pm$ 50 m measured from radar observations by \citet{Busch2011} results in a value of geometric albedo of 0.12 $\pm$ 0.04, which is slightly outside the limit of what has traditionally been considered primitive. This is also the case for (2) Pallas and the Pallas collisional family \citep[see, for example,][]{Ali-Lagoa2013}. 

In this work we apply an asteroid thermophysical model to fit \EV's outstandingly large set of infrared data obtained by NASA's Wide-field Infrared Survey Explorer (WISE) to derive its thermal inertia and draw conclusions about the characeteristic particle size of the regolith on its surface.

\section{Data}
\label{sec:data}

A general introduction to WISE can be found in \citet{Wright2010} and references therein. 
The NEOWISE project enhanced the WISE data processing system to allow detection and archiving of solar system objects \citep[for details, see][]{Mainzer2011a}. 
We obtain the data from the WISE All-Sky Single Exposure L1b Working Database, available via the IRSA/IPAC archive\footnote{http://irsa.ipac.caltech.edu/Missions/wise.html}.

WISE used four broad-band filters with isophotal wavelengths at 3.4, 4.6, 12, and 22 $\mu$m, referred to as W1, W2, W3, and W4, respectively \citep{Wright2010}.
As explained in \citet{Ali-Lagoa2013}, we follow a combination of criteria found in \citet{Mainzer2011b}, \citet{Masiero2011}, and \citet{Grav2012} to ensure the reliability of the data. 
We implement the correction to the red and blue calibrator discrepancy in W3 and W4, and we use a cone search radius of 0.3\arcsec\, centred on the MPC ephemeris of the object in our queries. 
All artefact flags other than p, P, and 0 and quality flags other than A, B, and C\footnote{Indicating signal-to-noise ratios $S/N > 10$, $3 > S/N > 10$, and $2> S/N >3$, respectively} are rejected, and we require the modified Julian date to be within four seconds of the time specified by the MPC. 
We ensure that the data is not contaminated by inertial sources by removing those points that return a positive match from the WISE Source Catalog within 6\arcsec. 
Finally, all remaining observations in band W1 were rejected since they are fewer than 40\% of the data in the band with the maximum number of detections, namely W3. 
These criteria give a total of 489 useful data points, 158 in W2, 190 in W3, and 141 in W4. 

Between \EV's first and last observations by WISE, taken in 2010 Jan 25 and March 7, the asteroid heliocentric distance decreased slightly, from 1.043 AU to 1.028 AU, it drew closer to the Earth by $\sim$ 0.06 AU, from $\Delta = 0.335$ AU to $\Delta = 0.273$ AU, and the phase angle increased from 71.5\dg to 75.3\dg.

\section{Thermophysical modelling of \EV}
\label{sec:TPM}

In this section we briefly describe the most relevant aspects of the thermophysical model (TPM) we employ. 
For more details, see \citet{Delbo2007}, \citet{Delbo2009}, and \citet{Mueller2007}. 
The technique consists of modelling the observed flux as a function of a given set of parameters and finding the set of parameter values, in our case thermal inertia, surface roughness and a scale factor $s$ for the asteroid shape, that minimise the $\chi^2$, i.e., 
\begin{equation} 
\label{eq:chi2} 
\chi^2 = \sum_i \frac{\left(s^2F_i - f_i\right)^2}{\sigma_i^2}, 
\end{equation} 
where $i$ runs through all observations, $s^2F_i$ is the model flux, $f_i$ is the measured flux, and $\sigma_i$ its corresponding error. 
$F_i$ is the unscaled mesh's model flux, which depends on its shape and spin axis orientation as a function of the geometry of the observation --phase angle and heliocentric and geocentric distances-- and the asteroid's albedo, thermal inertia, and macroscopic surface roughness, the last three assumed to be constant in time and throughout the surface. 
We also assume that thermal inertia does not depend on the temperature. 
The factor $s^2$ is related to how we model the size of the object. 
Each vertex of mesh is characterized by a vector in a given reference frame whose modulus is expressed in some given units. 
By multiplying all these vectors' moduli by the same linear scale $s$, we are able to change the model's size, and this factor is left to vary free and adjusted to minimise the $\chi^2$. 
But because the model flux depends on the object's area projected towards the observer, which depends on the square of this scaling factor,  we have that the model flux is $s^2F_i$.

The shape of the asteroid is represented by a set of 512 triangular facets based on a detailed radar shape obtained by \citep{Busch2011}. 
We also took the following physical properties as input for the model: rotation period $P = 3.725\pm 0.001$ h \citep{Galad2009}, absolute magnitude $H=20.0\pm 0.4$ (taken from the Small-Body Database of the Jet Propulsion Laboratory and rounded up to have one significant figure on the errorbar), and the retrograde pole-orientation solution preferred by \citet{Busch2011}, namely (180\dg, -84\dg) $\pm 10$\dg. 
Though \citet{Busch2011} conclude that \EV\, rotates retrograde, we also modelled the prograde solution and our analysis consistently favours the retrograde case (for more details, see Sect. \ref{sec:prograde}). 

In the absence of macroscopic surface roughness and zero thermal inertia, the temperature of each facet is a function of the incident solar radiation absorbed, which in turn depends on its albedo, the heliocentric distance, and the projection of its area onto a plane perpendicular to the direction towards the Sun. 
Surface roughness, is modelled by adding to each facet a hemispherical crater of opening angle $\gamma_{c}$ and crater surface density $\rho_{c}$, which is the ratio of the area of the craters to the area of the facets. 
Each crater is in turn divided into facets (typically $\sim$ 40) in order to introduce the effects of multiple scattering, which increases the surface temperature relative to the single scattering case. 
Non-illuminated crater facets (shadowed) may be heated by reflected sunlight and/or emission from other crater facets, so they also contribute to the flux if they are visible to the observer. 
The effects of thermal conduction towards layers beneath the surface are accounted for by numerically integrating the one-dimensional heat-diffusion equation in each crater facet for a given value of thermal inertia. 
The net energy absorbed is re-emitted assuming the facets emit like grey bodies ($\epsilon =$ 0.9). 
Our TPM does not account for shape shadowing effects, but we expect these to be negligible. To qualitatively justify this, we calculated the number of facets $n_b$ with potential blockers located $\ge$20\dg above their horizons. For the three shape models we used, namely the model introduced by \citet{Busch2011} with $n_f=3996$ facets and the simplified models with $n_f=1024$ and $n_f=512$, we find that $n_b/n_f$ is about 0.05, a very small percentage considering that having such potential blockers above the local horizon does not necessarily imply shadowing, since the sun can still be in a direction where it is visible and it would probably not be blocked simultaneously for all these facets. This conclusion can also be drawn from the small total view factor calculated for \EV\, by \citet{Rozitis2013}, which is the mean fraction of sky subtended by other facets of the model for any given facet. 

We thus calculated model fluxes for a wide range of preset values of thermal inertia, surface roughness, bolometric Bond albedo, and rotational phase $\varphi_0$, and adopted as best solution the one with the minimum $\chi^2$. 
Thermal inertia values run from $\Gamma$ = 0 to 2500 \SIu, values typical of perfectly insulating material and basaltic rock \citep{Jakosky1986}. 
Following the procedure of \citet{Mueller2007}, we used four preset combinations of ($\gamma_c$, $\rho_c$) to model surface roughness: no roughness (0\dg, 0), low roughness (45\dg, 0.5), medium roughness (68\dg, 0.8), and high roughness (90\dg, 0.9).   
As the rotational phase at the time of WISE observations cannot be accurately predicted from the rotational phase determined at the time of the radar observations (an error of 0.001 h for a period of about 3.725 h gives an error on the rotational phase of about 230 \dg/year), we treated $\varphi_0$ as a free parameter which took on all values multiple of 10\dg between 0\dg\ and 360\dg. For more details, see Sect. \ref{sec:rotphase} and \citet{Matter2011}, Finally, the bolometric Bond albedo was varied from 0.01 to 0.10.


\section{Results \label{sec:results}}
\label{sec:results}

In this section we present the best-fitting values of parameters, whereas in the following subsections we provide a detailed report on thermophysical analysis of \EV's WISE data, including a brief account of the discarded prograde rotational solution of \EV, the effect of simplifying the shape model to the one we used with a smaller number of facets, the relative minima found in $\chi^2$-$\Gamma$ space, and other possible sources of error that may explain the slightly high minimum effective $\chi^2$ achieved in our best-fit solution.

\begin{figure}  
\includegraphics[width=0.77\columnwidth]{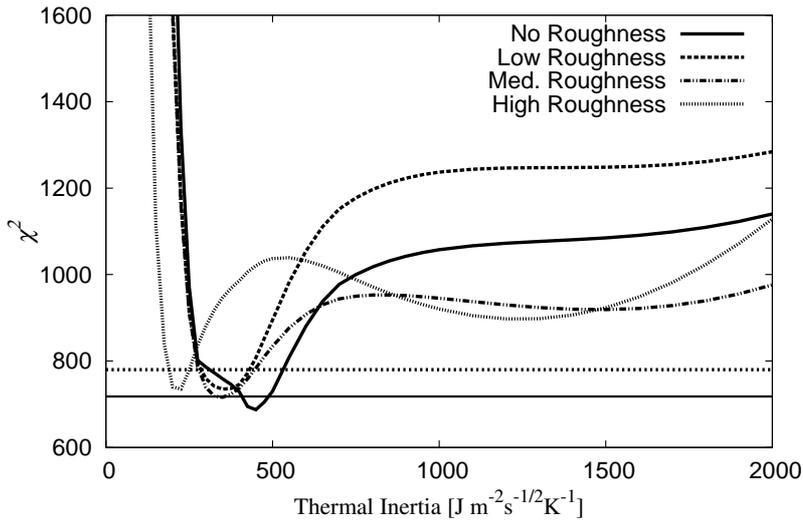}
\caption{
  $\chi^2$ versus thermal inertia for the runs with $\varphi_0 = 130$\dg. The smooth surface case  verifies the minimum $\chi^2$ of all runs. 
  The horizontal lines show the 1-$\sigma$ (black line) and 3-$\sigma$ (short-dashed line) levels at $\chi^2$ = 718 and $\chi^2$ = 780, respectively. 
}
\label{fig:chi2phi130}
\end{figure}

The minimum $\chi^2$ corresponds to an effective diameter $D = 368.9 \,\mathrm{m}$ (the diameter of the sphere with the same volume as the scaled shape model), thermal inertia $\Gamma = 450$ \SIu, zero roughness ($\gamma_c=0$, $\rho_c=0$), Bond albedo 0.08, and rotational phase $\varphi_0 = 130$\dg\ of the retrograde rotation model. 
In Fig. \ref{fig:chi2phi130}, we plot $\chi^2$ vs. thermal inertia for the $\varphi_0=130$\dg\ models. 
Interestingly, the $\chi^2$ curve presents more than one relative minima for some cases, a situation never reported before in the literature (for a more detailed discussion, see Sect. \ref{sec:facets}). 

Our minimum $\chi^2 \approx 687$ is of the order (though somewhat larger) of the effective number of degrees of freedom $\nu = N-n = 485$, where $N$ is the number of data points, and $n = 4$ is the number of free parameters. 
Assuming that the data errors are normally distributed, our $\chi^2$ statistic has a standard deviation of the order of $\sigma \sim \sqrt{2\nu} \simeq 31$ \citep[see, for example,][]{Press1986}. 
To give a better idea of the goodness-of-fit and to estimate an errorbar for our best-fit parameters, we include the 1-$\sigma$ and 3-$\sigma$ levels in Fig. \ref{fig:chi2phi130}, represented as a solid and dashed horizontal line, respectively. 
Within a 1-$\sigma$ confidence level, we can constrain $\Gamma$ to be within the interval (410, 490) \SIu, and the surface roughness at a macroscopic level to be negligible, though the minimum $\chi^2$ of the medium-roughness case is within the 1-$\sigma$ limit. 
On the other hand, at the 3-$\sigma$ level, we have $\Gamma \in (310,530)$ \SIu, and it is not possible to constrain roughness. 
The effective diameter is $D=370 \pm 20\,\mathrm{m}$ at the 3-$\sigma$ level and $D=368.9 \pm 0.5\,\mathrm{m}$ at the 1-$\sigma$ level, which are within the errorbar of previous estimates. However, these errorbars do not take into account the propagation of the error in the spin pole solution or the uncertainty in the Bond albedo, which we cannot constrain. Taking the conservatively broad range of values of Bond albedo, the 1-$\sigma$ errorbars in size and thermal inertia are increased: $D = $ 370 $\pm$ 6 m and $\Gamma = $ 450 $\pm$ 60 \SIu. The ensuing geometric visible albedo is $p_V = 0.13 \pm 0.05$, whose error is dominated by the uncertainty in the absolute magnitude. Note that this value of $p_V$ could be a small overestimate due to the systematic bias toward lower $H$-values (especially $H >$10) of widely used catalogues detected by \citet{Pravec2012}. 

\begin{figure}  
\includegraphics[width=0.49\columnwidth]{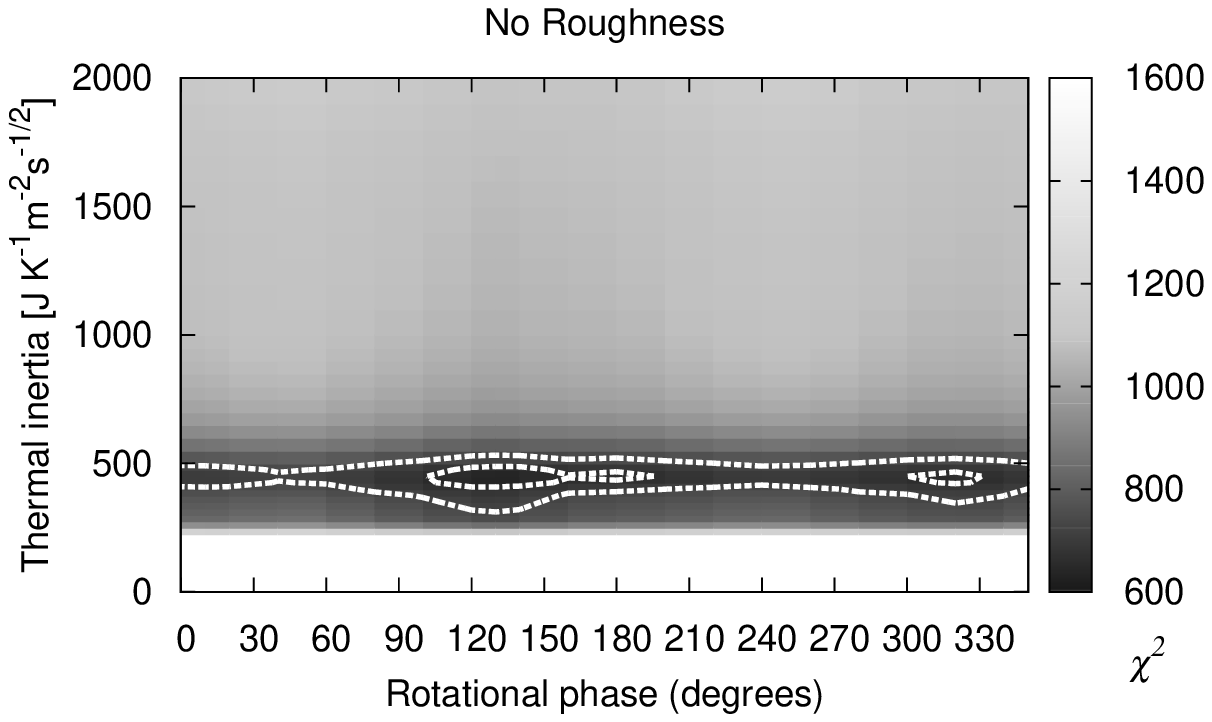}
\includegraphics[width=0.49\columnwidth]{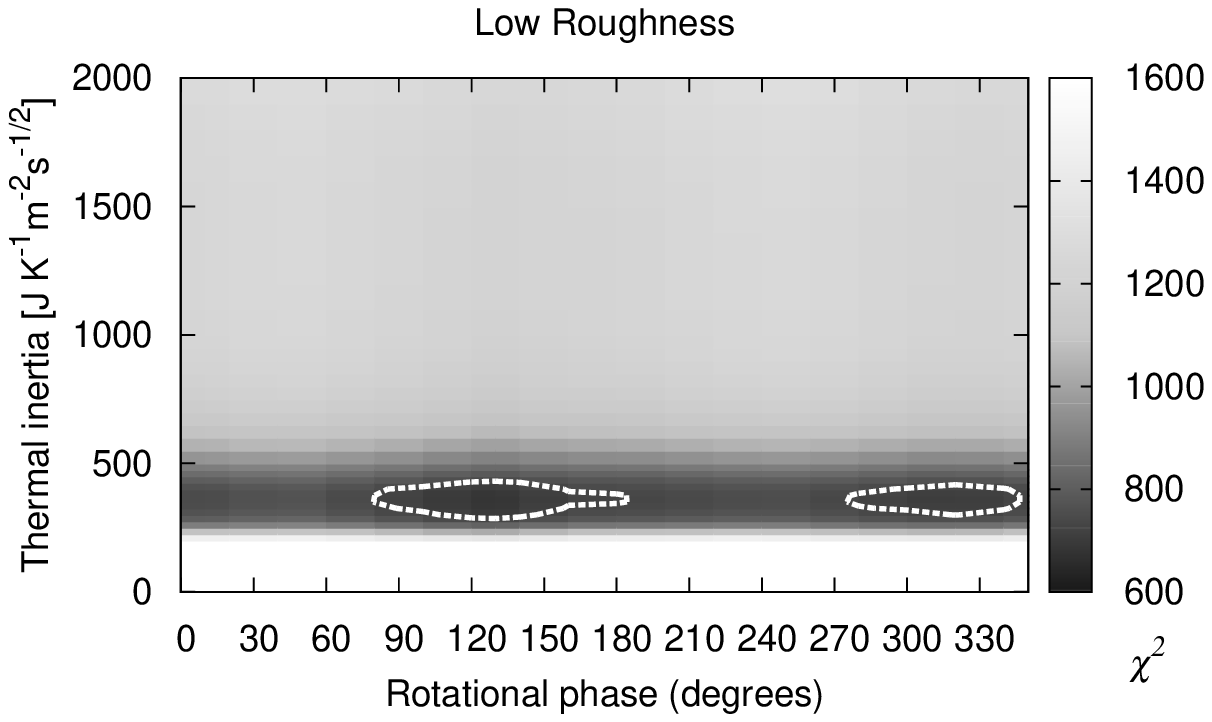}
\vspace{-1cm}

\includegraphics[width=0.49\columnwidth]{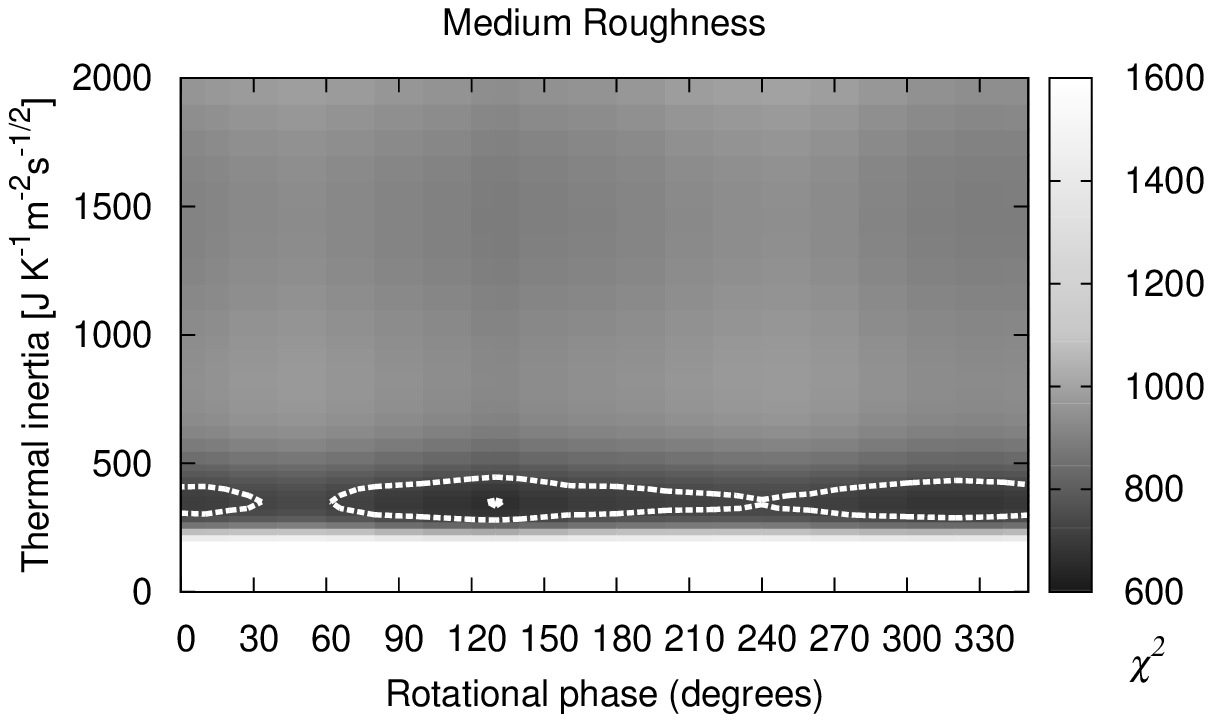}
\includegraphics[width=0.49\columnwidth]{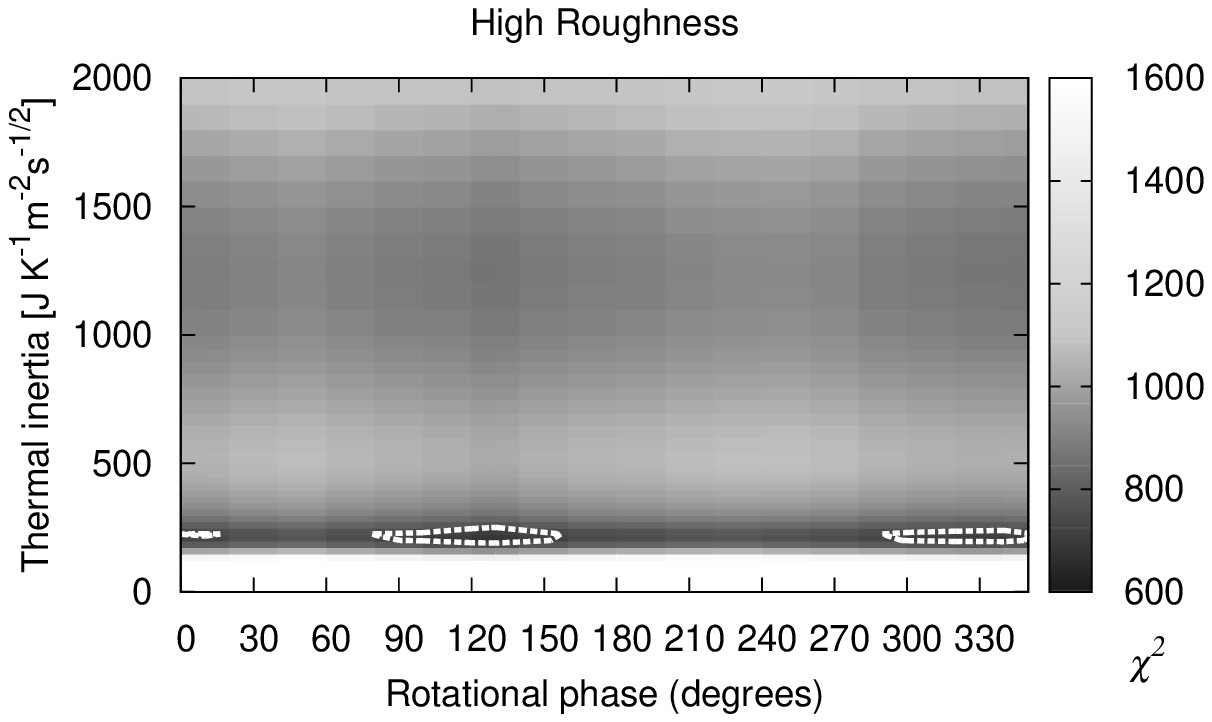}
\caption{
  Intensity map of $\chi^2$ values in $\Gamma$-$\varphi_0$ space. 
  The contour lines correspond to 1-$\sigma$ and 3-$\sigma$ values, the former only visible as the inner contours in the no-roughness and medium-roughness cases. 
}
\label{fig:chi2surf}
\end{figure}

The situation with the rotational phase is more complicated. 
In Fig. \ref{fig:chi2surf}, we show intensity maps of the $\chi^2$ values as functions of thermal inertia and rotational phase for the different roughness cases separately. 
The white, dashed lines show the 1-$\sigma$ and 3-$\sigma$ contours (when only one line is visible it corresponds to the latter case).
At a 1-$\sigma$ level of significance, the zero-roughness case presents two broad intervals of $\varphi_0$-values with minimum $\chi^2$, namely between 90\dg\ and 180\dg, and 300\dg and 330\dg.  
Again, we cannot constrain $\varphi_0$ at the 3-$\sigma$ level. 
In these plots, one can also see additional, shallower and broader minima at higher values of $\Gamma$ for the medium- and high-rougness models. 

\begin{figure}  
\includegraphics[width=0.77\columnwidth]{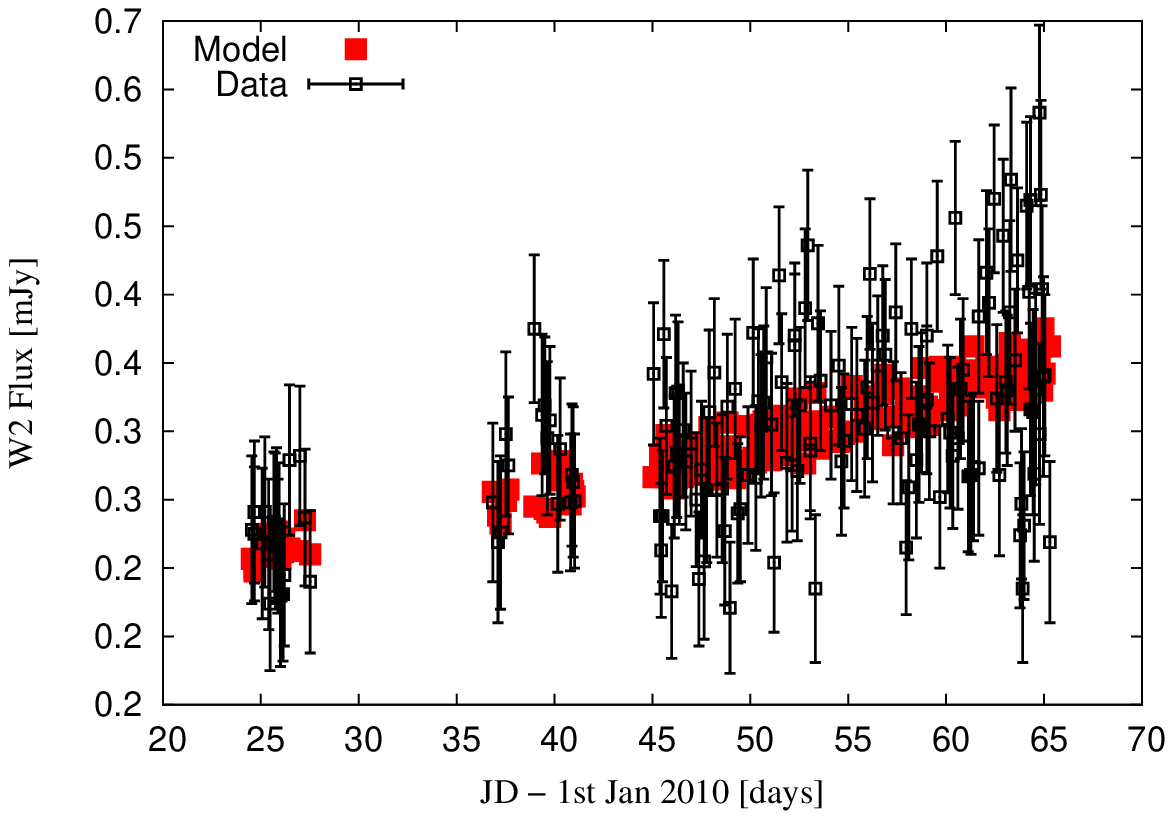}

\includegraphics[width=0.77\columnwidth]{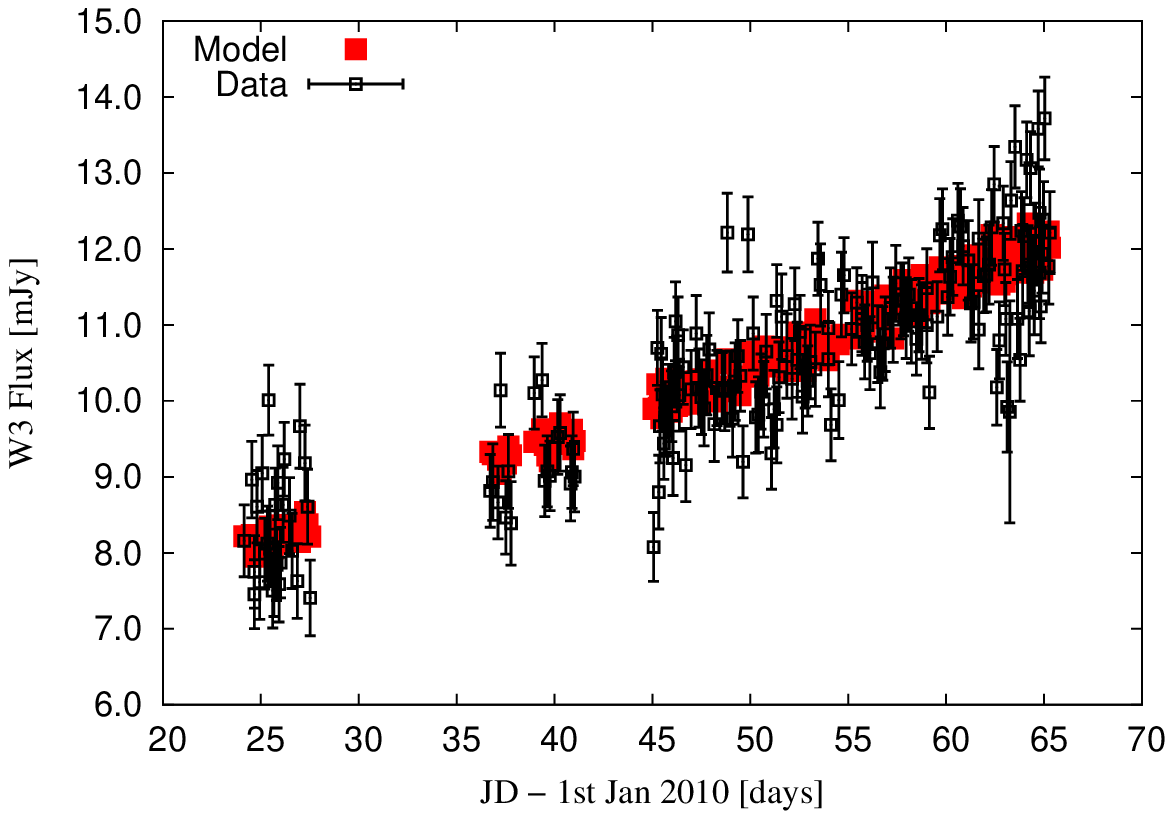}

\includegraphics[width=0.77\columnwidth]{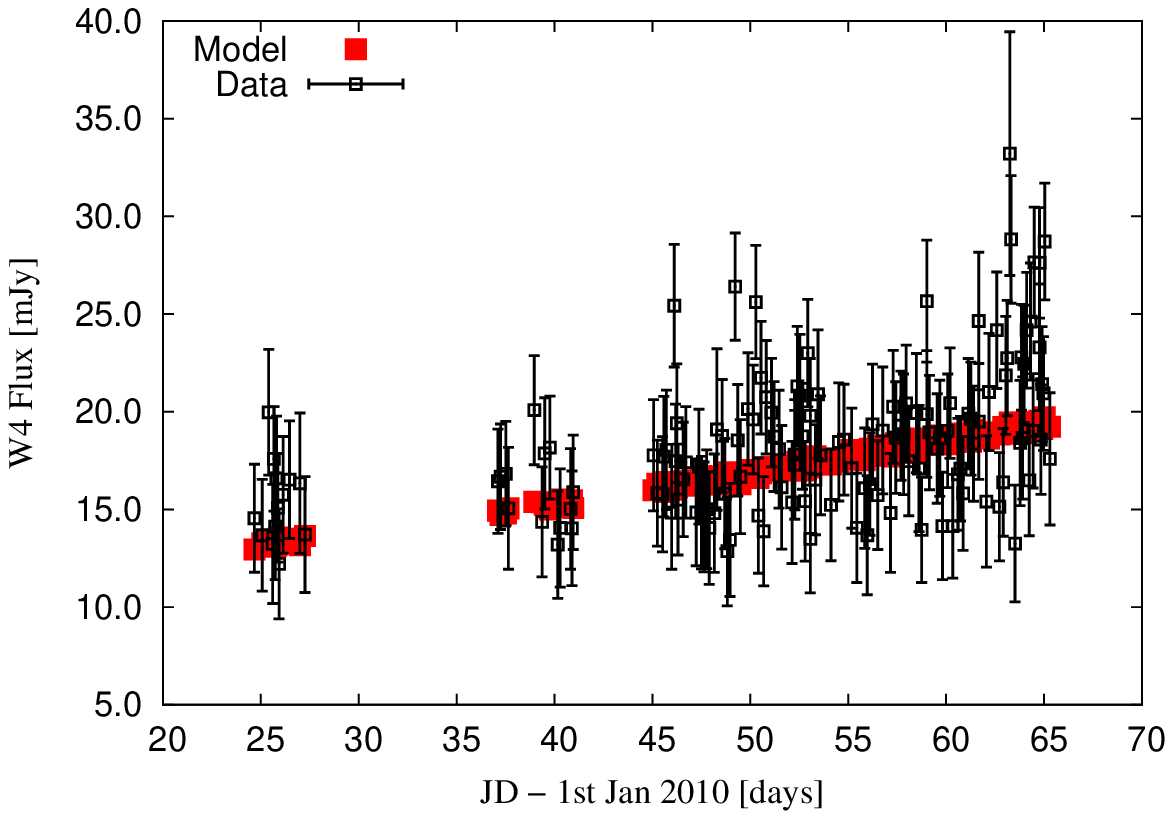} 

\caption{WISE band fluxes (black empty squares) and corresponding thermophysical model values (red filled squares) plotted versus date for the complete data set. The rotational phase is 130\dg at epoch 2455221.534375, i.e., the Julian Date (JD) on 2010 January 25 00:49:29.6 UT. 
}
\label{fig:modelfluxes}
\end{figure}

In Fig. \ref{fig:modelfluxes} we plot the observed and model fluxes versus Julian Date (hereafter JD, starting to count from 1st January 2010) for the model with best-fitting parameter values. 
The data show a large degree of scatter as compared to the model, and visual inspection of these plots does not help to choose among models with different but similar values of $\Gamma$ or surface roughness since the data flux variations in the different bands mostly look uncorrelated to each other and the model. 
But while effects of shape, spin pole orientation, and rotational period are smaller than the data scatter, the thermal inertia and the diameter are well constrained by the TPM, which produces much better results for \EV\, than the near-Earth asteroid thermal model \citep[NEATM, ][]{Harris1998}. NEATM considers a non-rotating, perfectly-difussing (Lambertian), zero-thermal-inertia, spherical asteroid, whereas asteroids surfaces are almost always non-spherical, non-Lambertian, not exactly perfect insulators, and may be macroscopically rough. Thus, this idealised model uses a free parameter $\eta$, the so-called ``infrared beaming parameter'', that modifies its surface temperature distribution to better fit the real thermal fluxes of asteroids. 
Now, the results reported for \EV\, in Table 1 of \citet{Masiero2011} cannot be directly compared because these authors do not fit the asteroid diameters when values are available from other more direct measurements (e.g., radar, stellar occultations), as is the case with \EV's radar size from \citet{Busch2011}. In addition, they use a different value of $H$ than we use here, namely 19.7. But from our NEATM, which essentially follows that described by \citet{Mainzer2011b} and \citet{Masiero2011} and which we validated in \citet{Ali-Lagoa2013}, we obtain higher but compatible values of infrared beaming parameter and size: $\eta = 2.2 \pm 0.4$ and 470 $\pm$ 70 m. 

\subsection{The prograde pole solution}
\label{sec:prograde}

\begin{figure}[h!]
  \centering
  \includegraphics[width=\columnwidth]{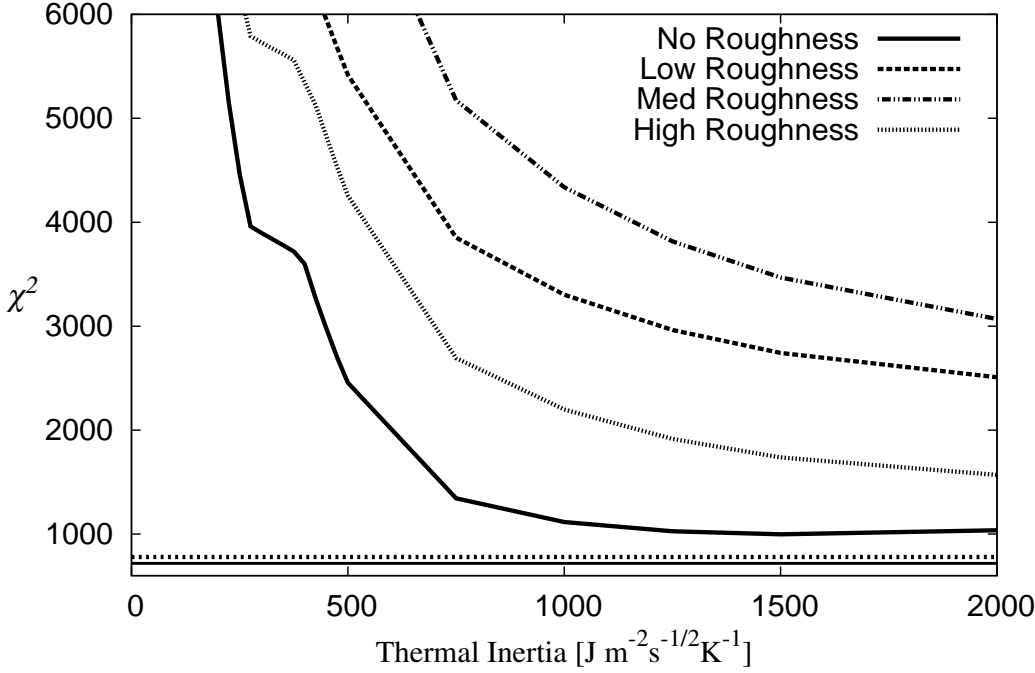}
  \caption{$\chi^2$ vs. thermal inertia for the prograde rotating model of \EV with four different values of surface roughness.
    The horizontal lines mark the value of the 1-$\sigma$ (continuous) and 3-$\sigma$ (short-dashed) levels of significance of the retrograde case (cf. Fig \ref{fig:chi2phi130}). 
    \label{fig:chi2prograde}
  }
\end{figure}

We modelled the two spin solutions of \EV\, given by \citet{Busch2011}. 
In Fig. \ref{fig:chi2prograde} we plot the $\chi^2$ vs. thermal inertia curves for the prograde case with four different values of roughness and rotational phase $\varphi_0 = 90^\circ$, which verifies the minimum $\chi^2$-values among all tested values of $\varphi_0$ (see Sect. \ref{sec:results}). 
The minimum $\chi^2$ for the prograde spin state clearly lies beyond the minimum the values of the 1-$\sigma$ and 3-$\sigma$ levels of significance of the retrograde case (black line and short-dashed line).  
These results are not unexpected since \citet{Busch2011} already concluded that \EV\, rotates retrograde, but they quickly illustrate how thermophysical modelling can help to discriminate between the two possible spin state solutions and at the same time confirm the radar results.  
  
\subsection{The effects of using a model with a reduced number of facets}
\label{sec:facets}

\begin{figure}[]
  \centering
  \includegraphics[width=\columnwidth]{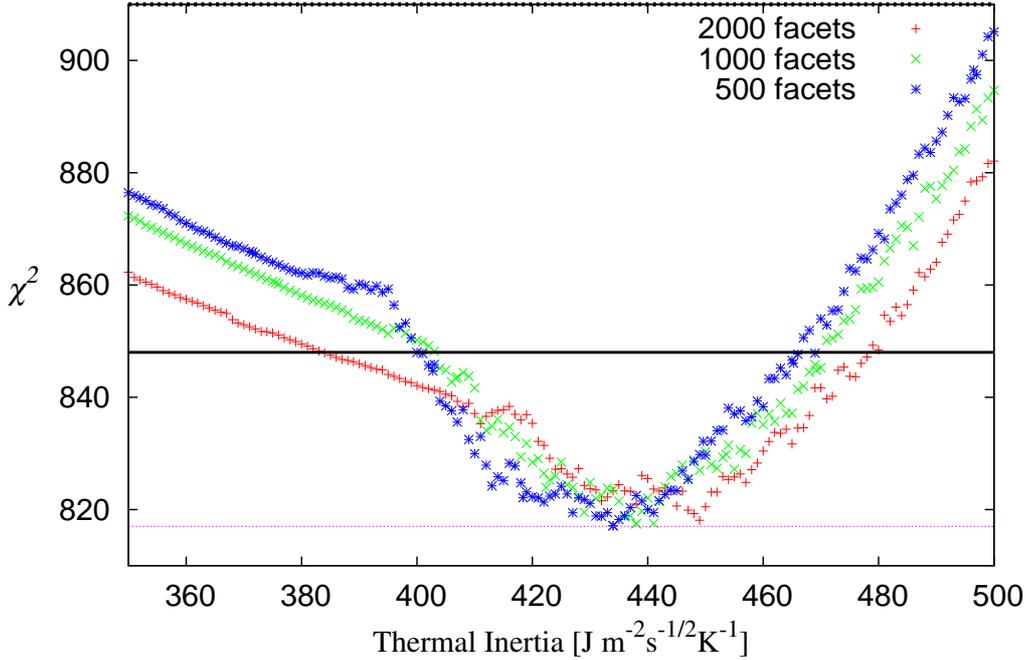} 
  \caption{$\chi^2$ vs. thermal inertia for three shape models with varying number of triangular facets. 
    The pink dashed horizontal line marks the minimum value of the $\chi^2$ in the figure, whereas the black continuous line marks the 3-$\sigma$ limit for this solution.  
    \label{fig:chi2facets}
  }
\end{figure}

The shape model we used resulted from the smoothing of the original shape model given by \citet{Busch2011}, i.e., it is a recomputed triangular mesh with the desired smaller number of triangular facets. 
While this will clearly reduce the computing time, which is  proportional to the number of facets, it may affect the best-fit values and introduce errors in the solution. 
To study the effects of this approximation, we carried out a sweep of thermal inertia values with a much narrower sampling step around the minimum for three shape models with different number of facets. 
In all three cases the best-fit solutions had $\varphi_0=130^\circ$ and zero roughness. 
As shown in Fig. \ref{fig:chi2facets}, where we plotted $\chi^2$ vs. thermal inertia for those models, the shift in the best-fit value of $\Gamma$ produced by this simplification is small compared with its estimated uncertainty. 
It is worth noting that the data set we used for this test is a preliminary one in which potential contamination by inertial sources was not addressed. 
In principle, very few non-inertial sources are expected to contribute significantly to the purely thermal bands W3 and W4, and in the particular case of \EV, the thermal emission in W2 is also expected to dominate (see Sect. \ref{sec:rlc} below). 
Indeed, none of our conclusions would be changed if we had not removed possible contamination from inertial sources, though the $\chi^2$ is reduced and the statistics improve slightly. 

On the other hand, these tests showed a puzzling feature: there are several relative minima in the $\chi^2$ curves of Fig. \ref{fig:chi2facets}. 
These features are unexpected since, in accordance with the long-proven validity of the model, the $\chi^2$ usually experiences a rapid increase when the values of the parameters are inadequate, and indeed such relative minima have never been reported before (to the best knowledge of the authors). 
One possibility is that these \emph{secondary} minima are caused by slightly different values of thermal inertia and/or roughness better fitting different subset of data, since those two parameters affect the shape of the lightcurve. 
Nevertheless, the data scatter is so large that it would be impossible to make a meaningful analysis by visual inspection, as may be done when the data set is much smaller, of the order of few tens of measurements. 

In addition, as already pointed out in the main text (see Figs. \ref{fig:chi2phi130} and \ref{fig:chi2surf}), there are also broad relative minima in $\chi^2$--$\Gamma$--$\varphi_0$ space for the non-zero roughness cases. 
These broad minima are narrower and deeper (in relative terms) as roughness increases, which may be due to a trade-off between thermal inertia and roughness: higher degrees of roughness will increase the surface temperatures, which may be partially compensated by high values of thermal inertia. 
However, as we increase the roughness to very high values, the thermal inertia values that may compensate it is increasingly smaller and hence the minima are narrower.

\subsection{Initial rotational phase as a free parameter}
\label{sec:rotphase}

We need to give the asteroid shape model the right orientation in space as a function of time. 
This is accomplished by applying an appropriate sequence of rotations to all vertices of the model at every step, as described, e.g., by \citet{Kaasalainen2002}.
These rotations are needed to perform the orientation as well as for changing from ecliptic coordinates in a reference frame centered at the asteroid to cartesian coordinates in a frame co-rotating with the asteroid. 
In particular, the orientation corresponding to the instantaneous rotational phase $\varphi$ is given by rotating the co-rotating frame an angle $\varphi$ about the $z$-axis, which is aligned with the spin axis of the asteroid, taken as input in ecliptic coordinates. Rotational phase at any given time $t$ is simply,
\begin{equation}
\varphi = \varphi_0 + 2\pi\frac{t-t_0}{P}
\end{equation}
where $\varphi_0$ is the corresponding value at $t_0$. An alternative way to motivate the use of an offset rotational phase $\varphi_0$ as a free parameter is given by \citet{Matter2011}. 
In essence, the error in the absolute rotational phase of an asteroid grows in time after the reference epoch at which its period is determined. 
From Eq. (3) of \citet{Matter2011} we obtain $\Delta_{\varphi} \ga $ 50\dg\, for \EV, which leads us to revise the value of this parameter. 

\begin{figure}[h!]
  \centering
  \includegraphics[width=\columnwidth]{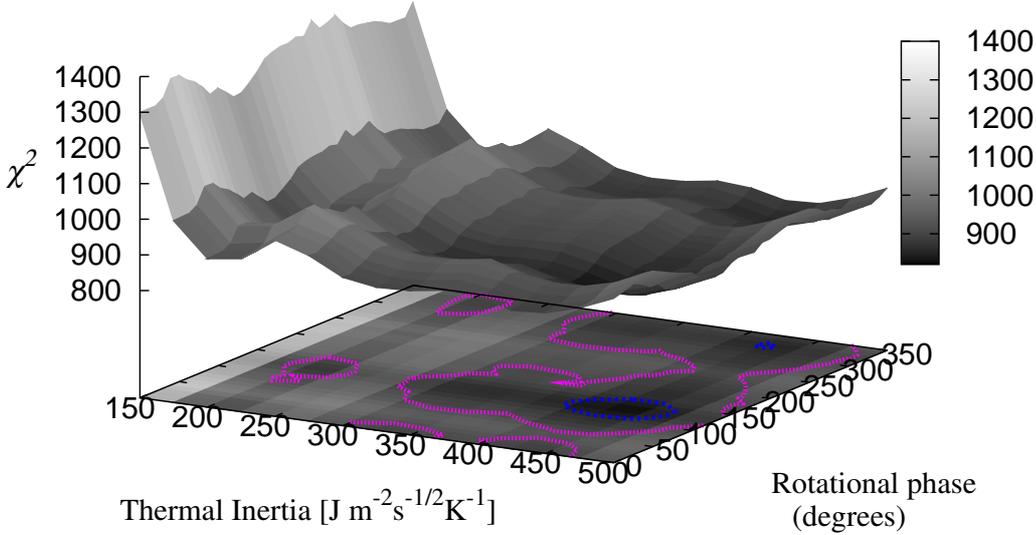}
  \caption{Minimum $\chi^2$ vs. thermal inertia vs. initial rotational phase. 
    ``Minimum $\chi^2$'' refers to the minimum among all four models with different surface roughness. 
    The projection onto the $\Gamma$-$\varphi_0$ plane shows the 1-$\sigma$ (dark blue) and 3-$\sigma$ (pink) contours. 
    The initial rotational phase that best fits the data is verified in epoch 2010 January 25 00:49:29.6 UT.
    \label{fig:chi2surface}
  }
\end{figure}

The thermophysical model does not, however, give a constraint for rotational phase at the 3-$\sigma$ level, but shows two intervals of possible $\varphi_0$-values at the 1-$\sigma$ level. 
In Fig. \ref{fig:chi2surface} we plot, for all pairs $\Gamma$--$\varphi_0$,  the value of the minimum $\chi^2$ of all four models with zero, low, medium and hig roughness for the data set with potential contamination from inertial sources. 
By showing the 1-$\sigma$ and 3-$\sigma$ countours, in dark-blue and pink colours, this plot illustrates at a glance the uncertainty intervals in the thermal inertia and rotational phase values for any value of roughness. 

\subsection{The reflected sunlight component in W2 data}
\label{sec:rlc}

The TPM used here does not include the reflected sunlight component of fluxes in bands W1 and W2 (W3 and W4 fluxes are thermal-emission dominated). 
Thoug we rejected \EV\ W1 data based on the minimum detection rate requirement (see Sect. \ref{sec:data}), we would not have used this band in our modelling since W1 is dominated by reflected sunlight. Using the NEATM solution with $D=0.4$ km, $\eta=2.0$ and $p_{IR}=0.10$, we estimate that $\sim 2/3$ of the flux would be reflected sunlight. On the other hand, this component may contribute $\sim$3\% to the total W2 flux of \EV. 
Failing to account for this could slightly but noticeably bias our results since, assuming\footnote{The largest SNRs of \EV's W2 data are $\sim$10, though most have lower values, so this would be an upper limit.} that all W2 points have signal-to-noise ratios (SNR) of 10, and given that we use 158 W2 points in our analysis, we would achieve an error $\sim$1\% on the mean W2 flux. 
Nevertheless, the two-parameter ($H$, $G$) phase function widely used to estimate the reflected sunlight for a given observation geometry \citep{Bowell1989} has an uncertainty of about 5\% in the final model flux in the W2 band. 
Because the contribution of the reflected light component itself is comparable to this uncertainty, we decided not to include it in the model. 
  
\subsection{The large concavity in \EV's shape model}

\begin{figure}  
\includegraphics[width=0.77\columnwidth]{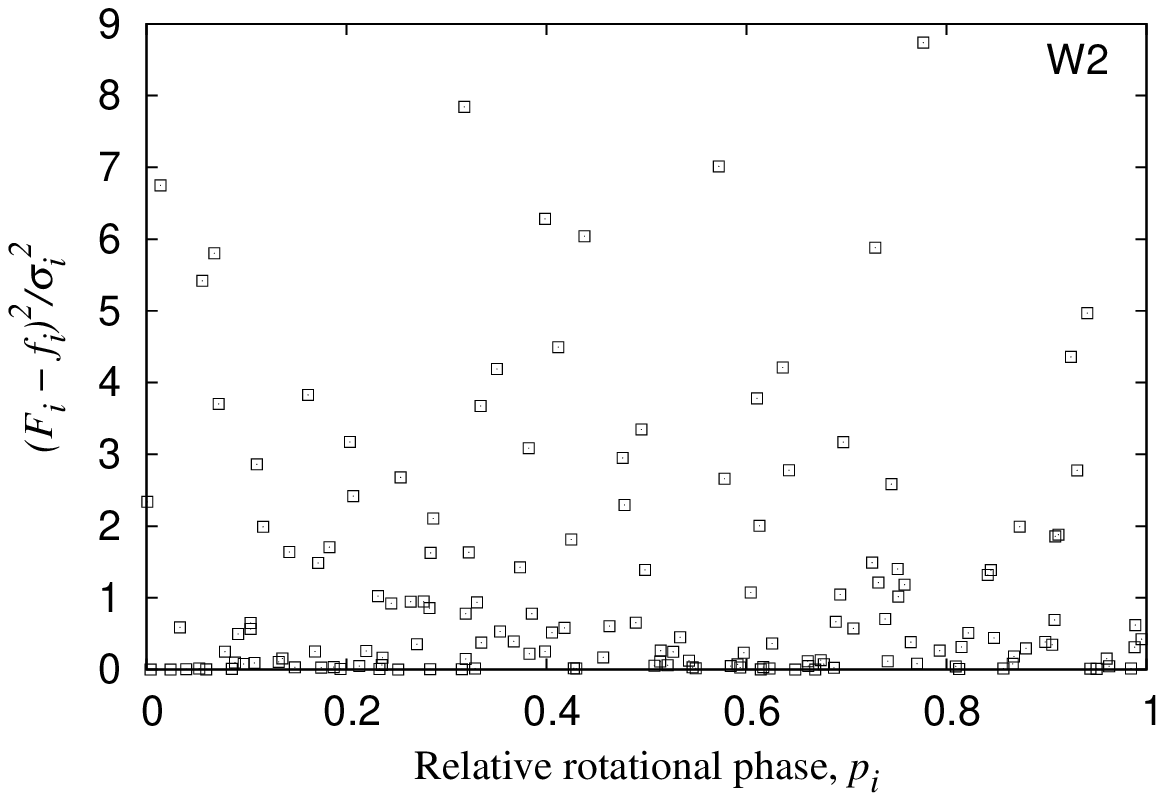}
\vspace{-0.5cm}

\includegraphics[width=0.77\columnwidth]{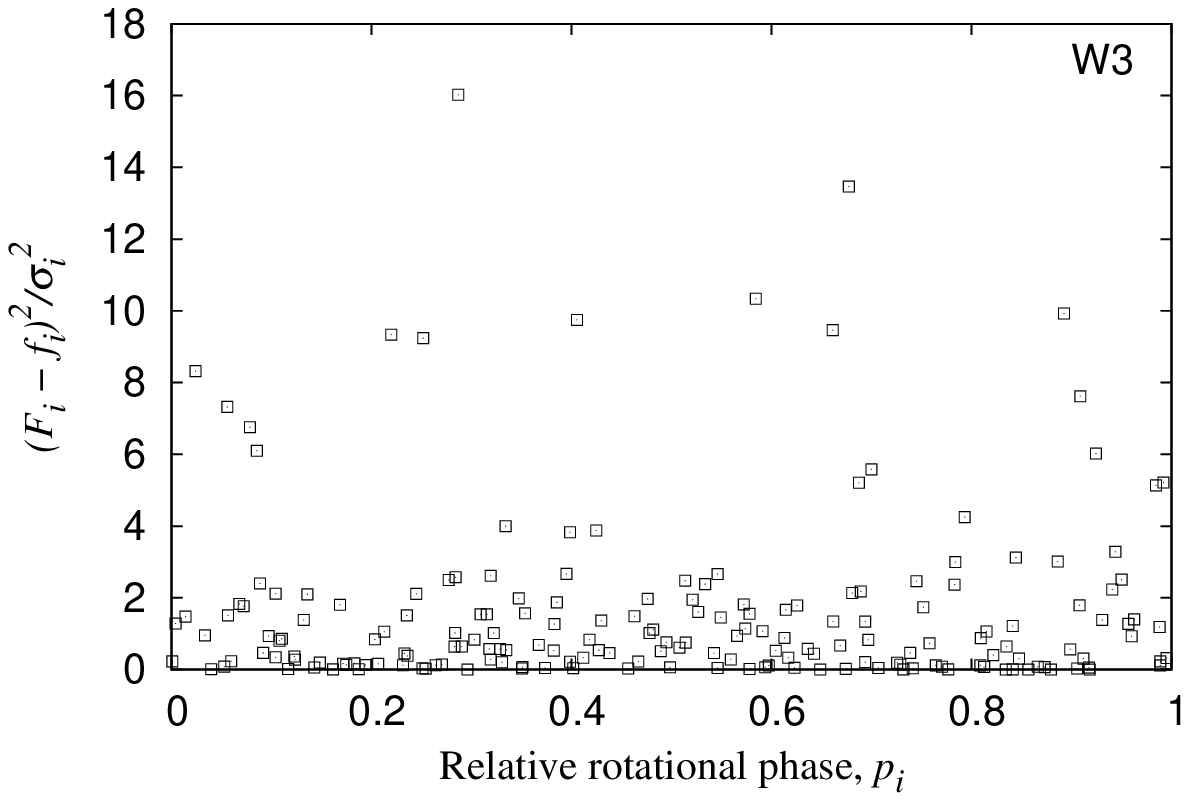}
\vspace{-0.5cm}
  
\includegraphics[width=0.77\columnwidth]{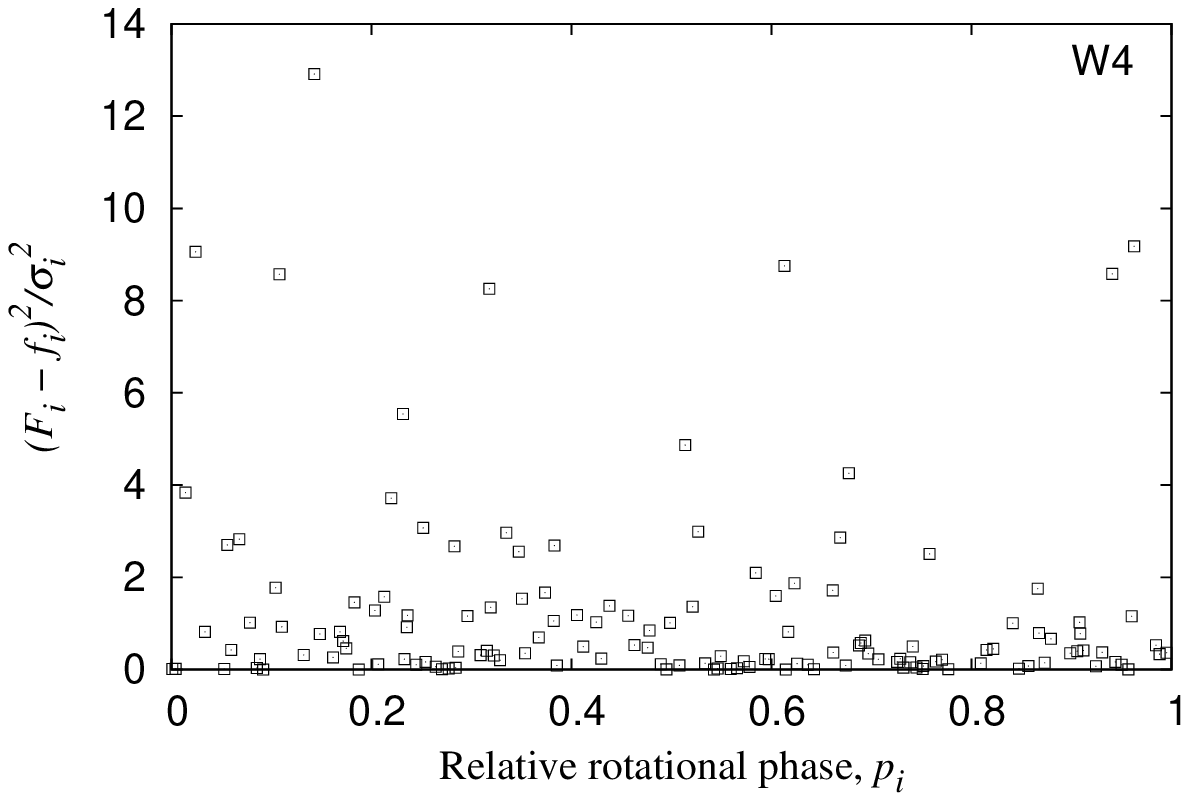}
\caption{
  Model--data deviations plotted against rotational phase for \EV's data in bands W2, W3 and W4. The rotational phase is calculated from Eq. \ref{eq:rotphase}
}
\label{fig:CheckConc}
\end{figure}

\citet{Busch2011} report that shape models without a large concavity could not fit \EV\, radar data. 
Though the concavity is still clearly visible in the simplified mesh, this feature could potentially be a source of inaccuracies given our simplification of the shape model and the multiple reflections between facets within the concavity. If this effect was significant, we would expect the model fluxes to deviate periodically in accordance with the phases of the period in which the concavity is visible to the observer. 
To test this, in Fig. \ref{fig:CheckConc} we plotted $\left((F_i - f_i)/\sigma_i\right)^2$ vs. relative rotational phase, $p_i$, for each WISE filer. Here, $F_i$ are the model fluxes and $f_i \pm \sigma_i$ are the data points and corresponding errors at epoch of observation, $t_i$. 
The relative rotational phase is calculated from the epoch of observation as follows:
\begin{equation} 
\label{eq:rotphase}
p_i = \frac{t_i - t_0}{P} - \textrm{floor}\left(\frac{t_i-t_0}{P}\right),
\end{equation}
where $P$ is the period and $t_0$ is the first epoch of observation in \EV's data set, which is arbitrarily chosen here as phase zero. 
In view of Fig. \ref{fig:CheckConc}, there seems to be no periodic maxima in the deviations of the model fluxes with respect to the data consistent throughout the three WISE bands, so the concavity's effect is either satisfactorily accounted for by the model or, more likely, its effects are smaller than the data fluctuations. 
In any case, this is consistent with the conclusions of \citet{Rozitis2013}, namely that the multiple reflections between its concavity and ensuing thermal emission do not have an influence on their prediction of \EV's Yarkovsky-O'Keefe-Radzievskii-Paddack (YORP) effect, therefore it would be unexpected to see an influence on its thermal emision.

\section{Regolith grain size}
Thermal inertia measurements can be used to determine the grain size of the surface regolith of \EV\ \citep{Gundlach2013}. In the following, we explain our strategy for the grain size determination.

\begin{figure}[h!!!!!]
  \centering
  
  \includegraphics[width=\columnwidth]{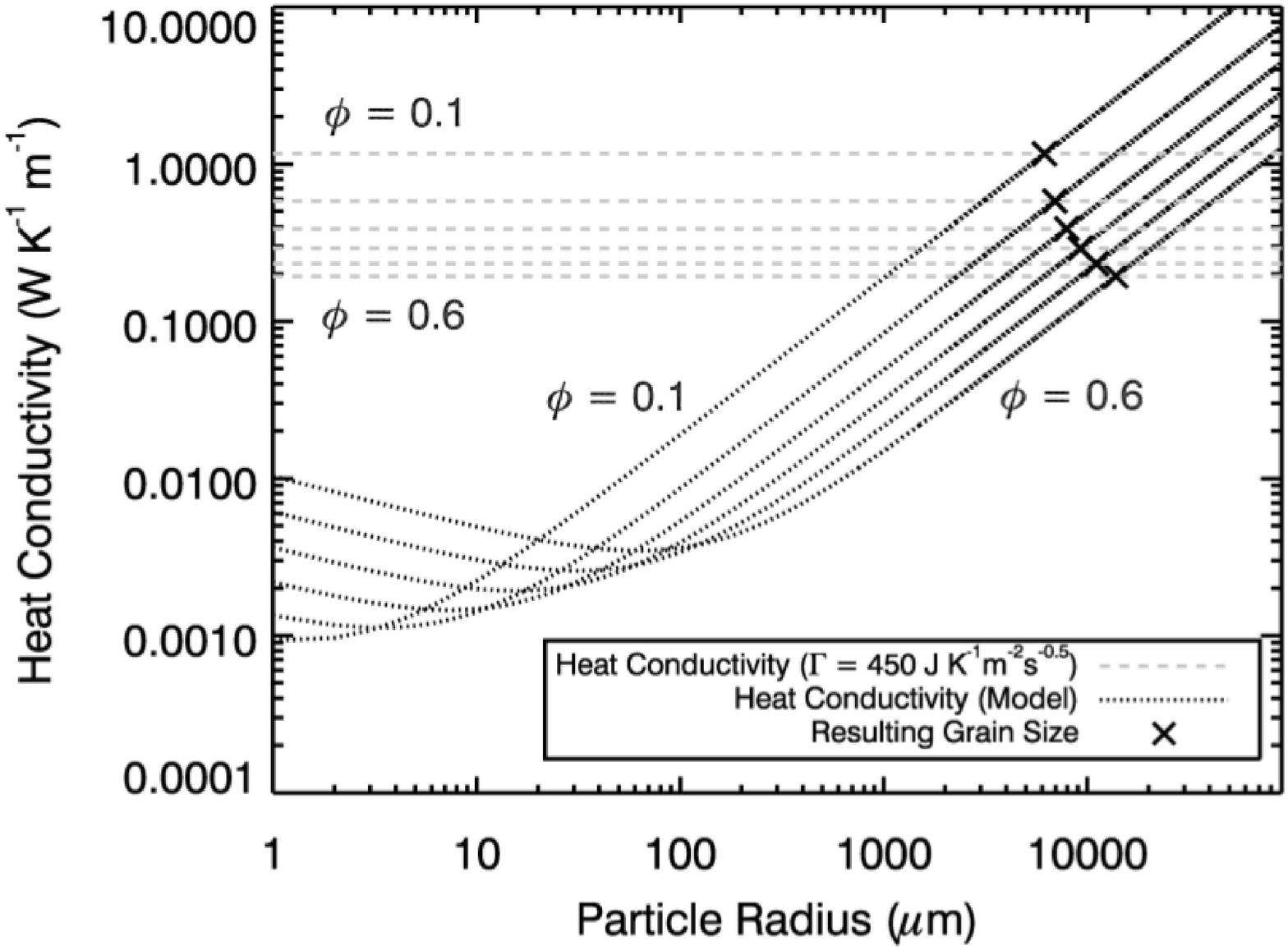}
  \caption{Grain-size analysis for the surface regolith of \EV. To estimate the mean grain size of the surface regolith, the heat conductivity of the surface material derived from the thermal inertia measurements (dashed lines) are compared with calculations of the heat conductivity of a model regolith (dotted curves) for six distinct volume filling factors $\phi = 0.1, \ldots 0.6$. }
  \label{FigGrainSize}
\end{figure}

First, the thermal inertia $\Gamma$ is used to derive the heat conductivity $\lambda$ of the surface regolith from
\begin{equation}
\lambda \, = \, \frac{\Gamma^2}{\phi \, \rho \, c}  \, \mathrm{,}
\label{basti1}
\end{equation}
assuming plausible values for the volume filling factor $\phi$, the mass density $\rho$, and the heat capacity $c$ of the surface regolith.
Since the packing density of the surface material is not known, the volume filling factor $\phi$ is treated as a free parameter and is varied between $\phi = 0.1$ (extremely fluffy packing, plausible only for small regolith particles and low gravitational accelerations) and $\phi = 0.6$ (close to the densest packing of equal-sized particles) in steps of $\Delta \phi = 0.1$. For the density and the heat capacity of the surface material, laboratory measurements of the density and heat conductivity of representative meteorites for C-type asteroids (CM2-type Cold Bokkeveld and CK4-type NWA 5515) are used, i.e., $\rho = 3110 \, \mathrm{kg m^{-3}}$ and $c = 560 \, \mathrm{J \, kg^{-1} \, K^{-1}}$ \citep{Opeil2010}. The dashed lines in Fig. \ref{FigGrainSize} show the derived values of the heat conductivity, following Eq. \ref{basti1}, for the different volume filling factors of the material and for a thermal-inertia value of 450 \SIu. 

To derive from the thermal-conductivity value a typical regolith-grain size, we calculate the heat conductivity of the surface material using a model for the heat conductivity of granular material in vacuum \citep[for details, see ][]{Gundlach2013}, which gives
\begin{equation}
\lambda(r, T, \phi) \, =  \,  \lambda_{\rm solid}(T) \, H(r,T,\phi) \, + \, 8 \, \sigma \, \epsilon \, T^3 \, \Lambda(r,\phi)  \, \mathrm{.}
\label{basti2}
\end{equation}
Here, $r$ and $T$ are the mean regolith-grain radius and the regolith temperature, respectively. The first term on the rhs. of Eq. \ref{basti2} describes the heat conduction through the solid network of regolith particles. Here, $\lambda_{\rm solid}(T)$ and $H(r,T,\phi)$ are the heat conductivity of the bulk material of the regolith and the Hertz factor, respectively. The bulk heat conductivity of the surface material is derived from laboratory measurements of representative meteorites for C-type asteroids \citep[Cold Bokkeveld and NWA 5515; see ][]{Opeil2010} by taking the porosities of the meteorites into account, $\lambda_{\rm solid} = (1.19 + 2.1 \times 10^{-3} \, T\, {\rm [K]}) \, \mathrm{W \, m^{-1}\, K^{-1}}$. The Hertz factor describes the reduced heat flux through the contacts between the regolith particles and depends on the mean radius of the regolith particles, on their temperature $T$, and on the volume filling factor of the surface material \citep[for details, refer to ][]{Gundlach2013}. The Hertz factor also takes into account the irregularity of regolith-particle shapes and has been calibrated with lunar regolith \citep{Gundlach2013}.
\par
The second term in Eq. \ref{basti2} takes the radiative heat conduction through the loose packing of regolith grains into account. Here, $\sigma$, $\epsilon$ and $\Lambda(r,\phi)$ are the Stefan-Boltzmann constant, the emissivity of the regolith grains (assumed to be $\epsilon = 0.9$), and the mean free path of the photons within the pore space of the regolith. The mean free path of the photons within the regolith pore space depends on the volume filling factor of the material and on the radius of the regolith grains and reads $\Lambda = 1.34 \frac{1-\phi}{\phi} r$ \citep[see][for more details]{Gundlach2013}. The model predictions of the heat conductivity of the surface material, following Eq. \ref{basti2}, are shown in Fig. \ref{FigGrainSize} (dotted curves) for different volume filling factors.

In the last step, the grain size of the surface regolith is derived from the comparison between the heat conductivity derived from the thermal-inertia measurements (dashed lines in Fig. \ref{FigGrainSize}) and the modelled heat conductivity as described above (dotted curves in Fig. \ref{FigGrainSize}). The intersections of the respective curves are denoted by the crosses in Fig. \ref{FigGrainSize}. One can see that the resulting grain radius is generally smaller for lower volume filing factors. For the low gravitational acceleration of \EV\ of $g = 7.5 \times 10^{-5} \, \mathrm{m \, s^{-2}}$ \citep[based on typical C-complex bulk densities of 1400 kg m$^{-3}$ given by ][]{Britt2002}, it is possible that the inter-particle forces prevent the collapse of the regolith to its densest packing so that we cannot exclude low volume filling factors of $\phi = 0.1, \ldots 0.2$. In this case, the mean particle radius is $r = 6.6^{+1.3}_{-1.3} \, \mathrm{mm}$. If, however, the inter-particle forces are negligible with respect to gravity, then packing densities of $\phi = 0.5, \ldots 0.6$ are expected so that the mean particle radius becomes $r = 12.5^{+2.7}_{-2.6} \, \mathrm{mm}$. The errors of the grain size estimations are then determined by quadratically adding the error of the thermal inertia measurements and the error due to the uncertainty in volume filling factor. For the error of the thermal inertia measurements we used the lower limit of $\Gamma = 410 \, \mathrm{J \, m^2 \, s^{-1/2} \, K^{-1}}$ and an upper limit of $\Gamma = 490 \, \mathrm{J \, m^2 \, s^{-1/2} \, K^{-1}}$.

\section{Discussion}

The thermal inertia of \EV\ obtained here is higher than the average value of km-sized NEAs \citep[200 $\pm$ 40 \SIu,][]{Delbo2007}, something to be expected for a smaller object. When compared to equally-sized, S-type Itokawa \citep[750$^{+50}_{-300}$ \SIu, ][]{Mueller2005}, \EV's smaller thermal inertia may be explained by their different compositions (assuming similar degrees of porosity), since more primitive C-complex asteroid such as \EV have much lower metallic content (or indeed none) than ollivine/pyroxene-rich S-types surfaces. When compared with other two similarly-sized, C-complex asteroids, (162173) 1999 JU$_3$ and (175706) 1996 FG$_3$, \citep[for quick reference, see Table 3 of][and references therein]{Gundlach2013}, FG$_3$ is approximately a factor of 3 larger and its thermal inertia is a third of \EV's \citep{Wolters2011}, which result in its much smaller grain size of $0.03$--$0.2$ mm. On the other hand, \EV\, and JU$_3$ have comparable gravitational acceleration and thermal inertia \citep{Hasegawa2008} and their characteristic grain sizes are compatible within errorbars, though for \EV\ the maximum value is about a factor of 2 smaller. With its somewhat smaller grain size, \EV's thermal inertia matching that of JU$_3$ could indicate a higher thermal conductivity and would be consistent with the suggestion of \citet{Busch2011} from radar albedo measurements--which are above the average for the 17 C-class, near-Earth asteroids observed with radar-- that \EV's surface material may contain some amount of metal. 
This can put a constraint on the possible meteorite analogues for \EV. 
\citet{Reddy2012ev5} studied visible-to-near-infrared spectra of \EV\, and preferred almost non-metallic, low albedo CI chondrites based on spectral slope and an unconfirmed absorption band at 0.48 $\mu$m, but rejected as possible matches other carbonaceous chondrites (CR, CO, CH, and CK meteorites)  with similar spectral slopes but higher metal content.

\section{Conclusions}

In this work we have performed thermophysical modelling of WISE data of (341843) 2008 \EV\, using the two spin-pole solutions given by \citet{Busch2011}. 
Our results favour the retrograde case, in consistency with the conclusions of \citet{Busch2011}. 
The best-fit value of thermal inertia within 1-$\sigma$, i.e., $\Gamma = 450 \pm 60$ \SIu, is attained for a rotational phase of $\varphi_0=$ 130\dg$_{-30^\circ}^{+50^\circ}$ and considering no surface macroscopic roughness, though the last two parameters are not constrained at a 3-$\sigma$ level of significance. 
The effective diameter and geometric visible albedo are $D = 370 \pm 6 \,\mathrm{m}$ and $p_V = 0.13 \pm 0.05$, also consistent with previous determinations. The errorbar in $D$ is at the 1-$\sigma$ level and does not take into account the uncertainty in the spin pole solution, so it is a minimum error estimate. 
From the mentioned value of $\Gamma$, the model of thermal conductivity by \citet{Gundlach2013} results in a mean regolith-grain radius of $r = 6.6^{+1.3}_{-1.3}$ mm for small values of the volume filling factor, and $r = 12.5^{+2.7}_{-2.6}$ mm for the highest packing densities. 

\begin{acknowledgements}
  We thank Dr. B. Rozitis for a thorough and constructive review that greatly improved this work. 
  We have benefitted from discussions with J. Hanu\v{s}. 
  VAL acknowledges support from the project AYA2011-29489-C03-02 (MEC, former Spanish Ministry of Education and Science). 
  This work was supported by the grant 11-BS56-008 (Shocks) of the Agence National de la Recherche (ANR) of France.
  Bastian Gundlach was supported by DFG under Grant Bl 298/19-1. 
  JL acknowledges support from the projects AYA2011-29489-C03-02 and AYA2012-39115-C03-03 (MINECO, Spanish Ministry of Economy and Competitiveness). 
  This publication uses data products from NEOWISE, a project of the Jet Propulsion Laboratory/California Institute of Technology, funded by the Planetary Science Division of the NASA. 
  We made use of the NASA/ IPAC Infrared Science Archive, which is operated by the Jet Propulsion Laboratory, California Institute of Technology, under contract with the NASA.  
\end{acknowledgements}

\bibliographystyle{aa}
\bibliography{AsteroidsGeneral}

%
%
%
%
%

\end{document}